\documentclass[acmlarge]{acmart}
\makeatletter
\newcommand{\myconfshort}{\acmConference@shortname}
\newcommand{\myconffull}{\acmConference@name}
\newcommand{\myconfdate}{\acmConference@date}
\newcommand{\myconfloc}{\acmConference@venue}
\AtBeginDocument{
  \fancypagestyle{firstpagestyle}{
    \fancyhead{}%
    \fancyfoot[C]{}%
  }
  \fancyhf{}
  \fancyhead[LO]{\@headfootfont\shorttitle}%
  \fancyhead[RE]{\@headfootfont\@shortauthors}%
  \fancyhead[LE]{\@headfootfont\footnotesize \myconfshort, \myconfdate, \myconfloc}%
  \fancyhead[RO]{\@headfootfont\footnotesize \myconfshort, \myconfdate, \myconfloc}%
  \fancyfoot[C]{}%
}
\makeatother
\acmBooktitle{\conffull\@ (\confshort), \confdate, \confloc}

\copyrightyear{2026}
\acmYear{2026}
\setcopyright{cc}
\setcctype{by}
\acmConference[FAccT '26]{The 2026 ACM Conference on Fairness, Accountability, and Transparency}{June 25--28, 2026}{Montreal, QC, Canada}
\acmBooktitle{The 2026 ACM Conference on Fairness, Accountability, and Transparency (FAccT '26), June 25--28, 2026, Montreal, QC, Canada}
\acmDOI{10.1145/3805689.3812237}
\acmISBN{979-8-4007-2596-8/2026/06}

\graphicspath{{./images/}} 
\hypersetup{colorlinks=true, linkcolor=blue, urlcolor=blue, citecolor=blue}
\usepackage{hyperref}

\begin{document}

%%
%% The "title" command has an optional parameter,
%% allowing the author to define a "short title" to be used in page headers.

\title[Platforms’ Research API Data Access]{Platforms’ Research API Data Access: \newline
What Users See vs. What Researchers can Retrieve}
%\title[Auditing Meta and TikTok Research API Data Access]{Auditing Meta and TikTok Research API Data Access under Article~40(12) of the Digital Services Act}

%%
%% The "author" command and its associated commands are used to define
%% the authors and their affiliations.
%% Of note is the shared affiliation of the first two authors, and the
%% "authornote" and "authornotemark" commands
%% used to denote shared contribution to the research.
\author{Luka Bekavac}
\email{luka.bekavac@unisg.ch}
\orcid{0009-0009-3598-3012}
\affiliation{%
  \institution{University of St. Gallen}
  \city{St.Gallen}
  \country{Switzerland}
}

\author{Simon Mayer}
\email{simon.mayer@unisg.ch}
\orcid{0000-0001-6367-3454}
\affiliation{%
  \institution{University of St. Gallen}
  \city{St.Gallen}
  \country{Switzerland}
}

%%
%% By default, the full list of authors will be used in the page
%% headers. Often, this list is too long, and will overlap
%% other information printed in the page headers. This command allows
%% the author to define a more concise list
%% of authors' names for this purpose.
\renewcommand{\shortauthors}{Bekavac and Mayer}

%%
%% The abstract is a short summary of the work to be presented in the
%% article.
\begin{abstract}
Article 40(12) of the Digital Services Act (DSA) requires Very Large Online Platforms (VLOPs) to provide vetted researchers with access to publicly accessible data. While prior work has identified shortcomings of platform-provided data access mechanisms, existing research has not quantitatively assessed data quality and completeness in Research APIs across platforms, nor systematically mapped how exactly current access provisions fall short.
This paper presents a systematic audit of research-access modalities by comparing data obtained through platforms' Research APIs with data collected about the same platforms' user-visible public information environment (PIE). Focusing on two major platform APIs—the TikTok Research API and the Meta Content Library—we reconstruct the full information feeds for two controlled sockpuppet accounts during two election periods, and benchmark these against the data retrievable for the same posts through the corresponding Research APIs.
Our findings show systematic data loss through three classes of platform-imposed mechanisms: scope narrowing, metadata stripping, and operational restrictions. Together, these mechanisms implement overlapping filters that exclude large portions of the platforms PIE (up to~\textasciitilde50\%), strip essential contextual metadata (up to~\textasciitilde83\%), and impose severe technical constraints for researchers (down to~\textasciitilde1{,}000 requests per week). Viewed through a data quality lens, these filters primarily undermine the completeness of research-accessible data, resulting in a structurally biased representation. We conclude that, in their current form, Meta \& TikTok Research APIs fall short of supporting meaningful, independent auditing of systemic risks as envisioned under the DSA.

\end{abstract}

%%
%% The code below is generated by the tool at: http://dl.acm.org/ccs.cfm
%% Please copy and paste the code instead of the example below.
%%



%%
%% Keywords. The author(s) should pick words that accurately describe
%% the work being presented. Separate the keywords with commas.

%% The following are not a requirement, delete if not using
\received{20 February 2025}  %% inital submission date
\received[revised]{12 March 2025} %% interim new draft
\received[accepted]{5 June 2025}  %% publication version

%%
%% This command processes the author and affiliation and title
%% information and builds the first part of the formatted document.
\maketitle

\section{Introduction}
Large online platforms have become core communication infrastructures of contemporary societies, mediating how information is produced, distributed, and consumed at unprecedented scale. Social interaction, political discourse, news consumption, and economic activity increasingly take place within a small number of platform-mediated environments, whose algorithmic systems shape visibility, attention, and reach~\cite{vanDijck2018platformsociety,Plantin2018infrastructure}. As a result, understanding how these platforms operate---and how their design choices affect individuals and society---has become a central concern across computer science, the social sciences, law, and public policy.
Against this backdrop, access to data handled and generated by large online platforms enables increasingly important independent research \textit{into} platforms themselves. Such research could, for example, examine platform design, governance, and recommender systems, or investigate how technical and organizational choices shape information exposure, public discourse, and collective outcomes. From this perspective, access to platform data is not merely a methodological convenience, but a prerequisite for understanding and evaluating the systemic impacts of large online platforms on society.
Beyond investigation of platforms as societal core artifacts, a growing body of empirical research in other disciplines depends on platform data, spanning domains as diverse as public health and nutrition science~\cite{culotta2010towards}, marketing and consumer behavior~\cite{lambrecht2019algorithmic,SANTOS2022102457}, and Web architecture and information diffusion~\cite{10.1145/3511095.3536365,kleinberg2008viral}. In many of these cases, platform data are used to study broader social, economic, or informational phenomena, rather than the operation of the platforms themselves.
% "This access is going away, but it is/was/would be very valuable for society"
Over the past years, however, obtaining any platform data has become increasingly difficult as many platforms have restricted or shut down their public APIs, a development described as the “API-calypse”~\cite{Bruns19092019}. As a result, researchers have increasingly struggled to observe how information circulates online and how platforms and their recommender algorithms shape what users see~\cite{Bekavac-TORS-SOAP}. This does not only lead to siloization of societally valuable data with a small group of private companies (cf.~\cite{bekavac2026broken}), but also increasingly undermines the auditing of large online platforms itself.
% "Regulation is doing something against it"

In response, new regulatory approaches have emerged that aim to restore access to platform data and clarify what such access should comprise~\cite{Status_Report:_Mechanisms_for_researcher_access2024}. One of the most significant of these efforts is the European Union’s Digital Services Act (DSA)~\cite{DSA2022}, which establishes new transparency and data-access obligations for the most popular platforms involved in disseminating information between users and to the public.
Article 40(12) of the DSA requires Very Large Online Platforms (VLOPs) to provide vetted researchers with access to publicly accessible data in order to scrutinize systemic risks, including those affecting elections, civic discourse, and public security~\cite{DSA_Data_access_2025}. Platforms implement this obligation through a range of mechanisms, including granting researchers permission to scrape, providing data upon request and offering dedicated \textit{Research APIs}\cite{Mozilla_data_access_initiatives}. In practice, Research APIs have become the primary mechanism through which platforms operationalize Article~40(12), and they now constitute central instruments within the DSA’s broader accountability and auditing framework~\cite{Mozilla_data_access_initiatives,DSA_Data_access_2025}.
% "This does not work well"
%The DSA, however, does not define “publicly accessible data,” leaving platforms wide discretion over what information is made available to researchers.\cite{Keller2025PubliclyAccessible,Better_Access}
Existing scholarship however shows that current Research APIs provided by VLOPs face structural limitations across multiple domains, including access gatekeeping, technical and operational restrictions, data quality issues including coverage and metadata gaps, use and dissemination constraints, and legal and liability provisions~\cite{Jaursch_Ohme_Klinger_2024, Mozilla, mimizuka2025postpostapiagestudyingdigital}. Together, these constraints cast doubt on the practical usability of DSA-mandated access for independent scrutiny. Regulators have begun to acknowledge these problems as well: X was recently fined 120 million Euro for failing to provide sufficient access to researchers\footnote{\url{https://digital-strategy.ec.europa.eu/en/news/commission-fines-x-eu120-million-under-digital-services-act} (last accessed: 06.01.2026)}, and the European Commission has preliminarily found both TikTok and Meta in breach of their Article~40(12) obligations, citing burdensome procedures and tools that leave researchers with incomplete or unreliable (e.g., outdated) data.\footnote{\url{https://ec.europa.eu/commission/presscorner/detail/en/ip_25_2503} (last accessed: 12.01.2026)} These sanctions emphasize that the issues identified by scholars reflect structural shortcomings in how platforms operationalize the DSA’s transparency requirements.
% "However, we don't know how bad it is and where to adjust legislation"
What remains largely unknown, however, is the empirical magnitude and structure of these shortcomings. While recent audits have examined specific aspects of post availability and data decay over time~\cite{entrenaserrano2025tiktoksresearchapiproblems,Pearson17022025}, no study has systematically assessed how closely Research API outputs correspond to the public information environment that users actually encounter on platforms. In particular, we lack evidence on how much of this environment is visible through Research APIs, which categories of posts and metadata are excluded, and whether the resulting datasets allow researchers to observe the systemic risks that Article~40(12) is intended to make analyzable.
To address this gap, we conceptualize all data delivered by platforms to their users during use as the \textit{public information environment} (PIE). This includes content that is visible to all users—as also to non-logged-in visitors—as well as the associated metadata and contextual signals transmitted to users’ devices. The central question we investigate is whether researchers, relying on official Research APIs, can reconstruct the same PIE that platforms present to users.
In this paper, we answer this question through a quantitative audit that directly compares platform PIEs with the data retrievable via Research APIs for two VLOPs, TikTok and Instagram. Using controlled sockpuppet accounts, we reconstruct the complete set of posts, metadata, and contextual information delivered through TikTok’s \textit{For You} feed and Instagram’s \textit{Explore} feed to each sockpuppet during the 2024 U.S.\ presidential election or the 2025 German federal election. We then benchmark this user-facing PIE against the data accessible to vetted researchers through the platforms’ official Research APIs.
Overall, we show evidence of VLOP Research APIs providing a substantially incomplete and systematically biased representation of the PIE, which raises concerns about whether current Platforms implementations of Article 40(12) can meaningfully support independent scrutiny of systemic risks by independent researchers. Our specific contributions are:

\begin{itemize}
    \item We present an analysis methodology that uses sockpuppet accounts to reconstruct the PIE of posts and metadata delivered to sockpuppets on two major VLOPs, offering a baseline against which to evaluate DSA-mandated research access. The relevant code is provided as open source.

    \item We propose a categorization of mechanisms that lead to data loss in Research APIs. These emerge from platform-induced overlapping filters that constrain access through three main mechanisms: \textit{scope narrowing}, \textit{metadata stripping}, and \textit{operational restrictions}. 
    
    \item We conduct the first quantitative investigation of loss/non-provisioning of data from users' public information environment in the TikTok and Meta Research APIs. We thereby quantify what fraction of the public information environment is accessible through these platforms' Research APIs and demonstrate structured systematic bias in the available data that undermines systemic risk audits.

    \item Based on our methodological findings and the results of our sockpuppet-based investigation, we provide first proposals of how current DSA provisions could be modified based on the proposed categorization to strengthen the regulatory basis and enable meaningful systemic risk audits.
\end{itemize}

\section{Background and Related Work}

Article~40 of the DSA establishes a formal legal basis for independent researcher access to platform data, but its operationalization varies considerably across providers. In the absence of centralized or harmonized rules for processing access requests, each VLOP has developed its own procedures, criteria, and technical infrastructures~\cite{Status_Report:_Mechanisms_for_researcher_access2024}. Most platforms have introduced dedicated Research APIs or virtual data enclaves as the primary interfaces through which access to publicly accessible data is granted under Article~40(12)~\cite{Status_Report:_Mechanisms_for_researcher_access2024}. In practice most platforms require researchers to complete a formal application---and in many cases to accept extensive contractual terms---before access to the API is provided~\cite{Status_Report:_Mechanisms_for_researcher_access2024}.
A growing body of work documents significant challenges associated with these access pathways and an extensive documentation of these is available in the European Commission's \textit{Status Report: Mechanisms for researcher access to online platform data}~\cite{Status_Report:_Mechanisms_for_researcher_access2024}: Researchers report opaque and inconsistent vetting processes, unexplained application rejections, and delays that stretch over weeks or months~\cite{Status_Report:_Mechanisms_for_researcher_access2024,Jaursch_Ohme_Klinger_2024}. Platforms justify denials by adopting a narrow interpretation of Article~40(12), arguing that proposed research does not exclusively contribute to “the detection, identification, and understanding of systemic risks” within the EU~\cite{Status_Report:_Mechanisms_for_researcher_access2024}. Some platforms additionally require researchers to submit draft publications or analyses to the platform before release, raising concerns about independence and editorial influence~\cite{Status_Report:_Mechanisms_for_researcher_access2024}. Beyond these application-level barriers, prior literature identifies a set of structural constraints embedded in platforms’ terms of service and research-access agreements~\cite{Jaursch_Ohme_Klinger_2024,Mozilla,Keller2025PubliclyAccessible,mimizuka2025postpostapiagestudyingdigital, 2024-edmo}. These constraints fall into several categories:
\begin{itemize}
    \item \textbf{Access gatekeeping}, including restrictive eligibility criteria and narrow interpretations of systemic-risk relevance (cf.~\cite{Jaursch_Ohme_Klinger_2024}).
    \item \textbf{Technical and operational restrictions}, such as strict rate limits, unstable endpoints, scraping prohibitions, and the requirement to conduct analysis within restricted virtual data enclaves (cf.~\cite{Mozilla, 2024-edmo}).
    \item \textbf{Data coverage and metadata gaps}, where large portions of public content or contextual information are missing, incomplete, or inconsistently documented\footnote{\url{https://researchportal.vub.be/en/publications/tiktoks-research-api-problems-without-explanations/} (last accessed: 25.03.2026)} (cf.~\cite{Keller2025PubliclyAccessible, 2024-edmo}).
    \item \textbf{Use and dissemination constraints}, including onward-sharing prohibitions, export limitations, and requirements for pre-publication review (cf.~\cite{Status_Report:_Mechanisms_for_researcher_access2024, 2024-edmo}).
    \item \textbf{Legal and liability provisions} that impose heightened compliance burdens or risk on researchers (cf.~\cite{Keller2025PubliclyAccessible}).
\end{itemize}
Complementing this governance-oriented work, technical audits provide evidence of concrete data quality issues within platform-provided Research APIs. For TikTok, prior studies show missing public posts, inconsistent metadata, unexplainable account omissions, and discrepancies between live engagement metrics and those returned via the API~\cite{entrenaserrano2025tiktoksresearchapiproblems,Pearson17022025}. Similar assessments of Meta’s Content Library highlight gaps in account coverage, content-type exclusions, and significant differences between the parameters exposed to researchers and those transmitted to the browser during normal platform use~\cite{Status_Report:_Mechanisms_for_researcher_access2024, 2024-edmo}. Together, the literature indicates that the data made available under Article~40(12) diverges from what is visible to ordinary users.
A central reason for this discrepancy is that the DSA does not clearly define ``publicly accessible data.''~\cite{Keller2025PubliclyAccessible,Better_Access}. Platforms therefore retain discretion to decide which formats, account types, metadata fields, and historical records fall within the scope of Article~40(12). %Legal scholarship argues that, for systemic-risk research, ``publicly accessible'' could reasonably include (a) content visible to any logged-in user, (b) content surfaced by recommender systems such as TikTok \textit{For You} or Instagram \textit{Explore}, (c) content disseminated within large groups that function as de-facto public spaces, or (d) content that was visible at an earlier time but later deleted or archived~\cite{Keller2025PubliclyAccessible}.
The need for a unified definition of ``public data'' across platforms and how this could be balanced with user privacy protections is also made explicit in~\cite{Better_Access} and highlights why platform-defined notions of public accessibility often diverge from the information environment that researchers seek to study.
While prior work hence documents substantial procedural and technical obstacles to data access through benchmarks on or expert interviews\cite{2024-edmo}, it provides little evidence on their empirical consequences. In particular, we lack a systematic comparison between the public content that users actually see while using a platform and the content researchers are able to retrieve through official Article~40(12) interfaces. For the purposes of this study, and while acknowledging the above-mentioned debate on what is ``public data'', we define a platform's \textit{public information environment (PIE)}, as that content that originates from public accounts and is actively amplified to users beyond the poster’s direct network, making it in principle accessible to all users and, for several platforms, also to non-logged-in visitors.

% We therefore construct a ground-truth baseline of user-visible posts and metadata and benchmark it against the data available from platform Research APIs. The next section outlines our methodological approach.

\section{Methodology}

\begin{figure*}[t]
    \centering
    \includegraphics[width=\linewidth]{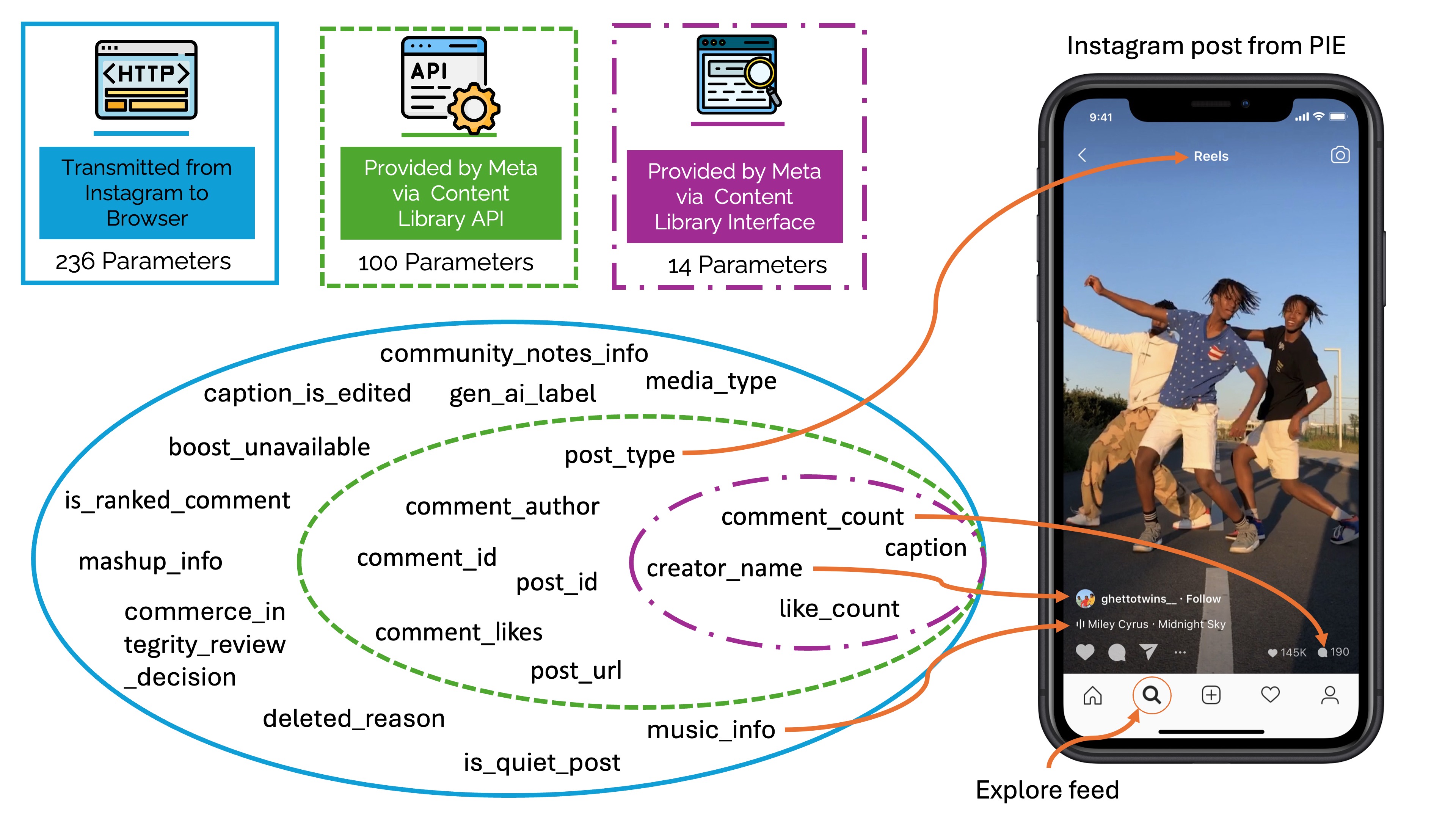}
    \caption{Data transmitted for a single Instagram post (excerpt): Comparison between parameters in full HTTP response (solid light blue, $n_{HTTP} = 236$), parameters that are accessible through the Meta Content Library API (dashed green, $n_{MCL-API} = 100$) and parameters that are accessible through the Meta Content Library User Interface (dot-dashed purple, $n_{MCL-UI} = 14$).}
    \Description{A diagram with three nested regions (a Venn-style 
hierarchy) showing parameters returned for a single Instagram post. 
The outermost solid blue region represents 236 parameters in the raw 
HTTP response from Instagram to the browser. Nested inside, a dashed 
green region represents the 100 parameters accessible via the Meta 
Content Library API. Nested inside that, a dot-dashed purple region 
represents the 14 parameters exposed through the Meta Content Library 
user interface. Sample parameter names are placed in their respective 
regions: the innermost UI-accessible set includes comment_count, 
creator_name, caption, and like_count; the API-only ring adds fields 
such as post_id, post_url, post_type, comment_id, comment_author, 
comment_likes, media_type, gen_ai_label, and community_notes_info; 
the HTTP-only outer ring adds fields such as is_ranked_comment, 
mashup_info, commerce_integrity_review_decision, deleted_reason, 
is_quiet_post, caption_is_edited, boost_unavailable, and music_info. 
On the right, a smartphone mockup shows an Instagram Reels post, with 
orange arrows connecting on-screen elements to the corresponding 
parameter names (Reels label to post_type, comment count to 
comment_count, account handle to creator_name, and audio track to 
music_info).}
    \label{fig:postdata}
\end{figure*}

To evaluate what fraction of the PIE is not visible through Article~40(12) mechanisms---and to investigate why it is not visible---we employ a design that allows for a direct, user-centric comparison between the content delivered to users and the content provided to researchers. In our methodology, we first construct a complete baseline of user-visible content using controlled sockpuppet accounts~\cite{Bekavac-TORS-SOAP}. We then query platforms' official Research APIs to obtain the ``research-accessible'' subset of this content. Finally, we compare the two obtained datasets to quantify the overall loss and to identify and investigate the responsible mechanisms.
Throughout this study, we treat as publicly accessible any post that is circulating in a platform's PIE and can be viewed with or without an account on the platform, as well as all metadata that is transmitted to the user agent (e.g., the browser) along with such posts.  To illustrate this while taking the Instagram platform as an example, Figure~\ref{fig:postdata} contrasts what parameters are sent as part of HTTP responses during interaction and what fraction of these are available via the Meta Content Library access modalities. Notably, we consider as publicly accessible all data that is delivered to the user’s device during platform use, even if certain elements are not visually rendered. For example, the primary key (PK-ID) of an Instagram post—which uniquely identifies the post within the platform’s internal systems. While not visible in the user interface of platforms, such data is nonetheless publicly accessible in the sense that it is delivered to any user and device viewing the post.

\subsection{Data Collection through Sockpuppet Accounts}

To collect the PIE displayed on the User Interfaces of platforms, we used the System for Observing and Analyzing Posts (SOAP)\footnote{\url{https://github.com/Interactions-HSG/SOAP}} (cf.~\cite{soap,Bekavac-TORS-SOAP}). SOAP enables controlled interaction with multiple social media platforms through sockpuppet accounts and captures the full data payload transmitted via platform API calls to the user’s browser.% Using this approach, we collected all posts recommended to a user, together with all metadata returned by the platform to the user’s device, including data that is not rendered in the user interface.
As illustrated in Figure~\ref{fig:postdata}, this includes all information transmitted via HTTP responses to the browser, regardless of whether it is displayed in the platform’s user interface. 
By using controlled sockpuppet accounts in combination with the SOAP system, we mitigate both privacy and data loss concerns in the data collection process. Privacy risks are reduced because no data from real individuals are collected as the sockpuppet accounts do not correspond to existing persons. Data loss is further minimized because all data are captured and stored at the moment a post is recommended to the user, rather than being collected retrospectively. This approach avoids common forms of ex post data loss associated with methods such as data donation or delayed scraping, where content may already have been deleted or become inaccessible by the time it is collected (e.g.,~\cite{Balogun_2025}).
We note that our sockpuppet-based collection methodology interacts with platform research terms in ways that require careful consideration. A discussion of the specific contractual clauses affected and the normative justification for our approach is provided in Appendix~\ref{app:legal-compliance}.
The sockpuppet accounts collected two complete PIE datasets covering several weeks of activity each: all posts recommended in Instagram’s \textit{Explore} feed from 16.07.2024 - 06.11.2024 during the U.S. presidential election period in 2024 and all posts recommended in TikTok’s \textit{For You} feed from 08.01.2025 - 25.02.2025 during the German federal elections in 2025. In total, this resulted in over 4000 posts collected and over 200 hours of sockpuppet-swiping. The sockpuppet accounts were configured to interact primarily with political content related to these elections, resulting in feeds containing election-related material. Importantly, both TikTok’s \textit{For You} feed and Instagram’s \textit{Explore} feeds recommend only content that is publicly accessible on the platform. Posts shown in these feeds hence originate from public accounts and are actively recommended to users beyond the poster’s direct network, making them in principle visible to all users; and, for some platforms, also to non-logged-in visitors. As such, the collected datasets represent content that platforms themselves deem suitable for public dissemination.
SOAP captures all of the content delivered to the user’s browser, in the exact order and contextual framing in which it was transmitted by platforms. As a result, it provides a complete baseline of the user-facing PIE in contexts directly relevant to systemic risk analysis under Article~34(1)(c) DSA, including potential negative effects on civic discourse, electoral processes, and public security. For a more detailed description of the set-up and deductive coding of political content we refer to Appendix \ref{app:Sockpuppets}.

\subsection{Data Collection through Research APIs}
\label{sec:data-collection-apis}

To obtain the research-accessible subset of the PIE, we collected data through the platforms’ official Research APIs: the Meta Content Library\footnote{\url{https://transparency.meta.com/researchtools/meta-content-library/} (last accessed: 13.01.2026)} and the TikTok Research API\footnote{\url{https://developers.tiktok.com/products/research-api/} (last accessed: 13.01.2026)}.
Access to both Research APIs was provided through controlled environments. Meta distinguishes between the \textit{Content Library Interface}, a web-based controlled-access environment intended for exploratory use, and the \textit{Content Library API}, which enables programmatic queries but must be used within a Virtual Data Enclave (VDE). TikTok similarly provides access to its Research API through a Virtual Compute Environment (VCE), in which data queries and analysis are conducted under platform-defined constraints. 
Comparing sockpuppet-collected posts with researcher-accessible data requires matching content across two systems that use different identifier schemes. For TikTok, deterministic matching is possible: the platform's native \texttt{video\_id}, contained in each post's URL, can be directly queried through the Research API, consistent with the methodology used in prior audit studies~\cite{entrenaserrano2025tiktoksresearchapiproblems, Pearson17022025}. For Meta, however, deterministic matching is not possible. The Meta Content Library generates privacy-preserving creator and post identifiers that differ from the platform's native identifiers, and does not support retrieval via platform URLs or native post IDs~\cite{2024-edmo}. We therefore implemented a multi-stage matching pipeline for Meta, combining automated bulk queries with manual verification for ambiguous cases. Coverage rates reflect the proportion of sockpuppet-scraped posts that were retrievable under each platform's API constraints at the time of querying. API queries were conducted at multiple points between March and June 2025 for both platforms. The full matching pipeline, including platform-specific procedures, edge-case handling, and replicability considerations, is described in Appendix~\ref{app:data-pipeline}.

\section{Results: Data Loss in Research APIs}
\label{sec:overlapping-filters}
We compared the PIE collected via the sockpuppet accounts with the data accessible to us through the two platforms' Research APIs. The platforms' respective definitions of “publicly accessible data” applied as per the time the data was collected. Notably, both Meta and TikTok have revised their interpretations of what counts as publicly accessible data, and therefore the data made available through their research access mechanisms, multiple times since our data collection. For an overview of these changes, see the Meta Changelog\footnote{\url{https://developers.facebook.com/docs/content-library-and-api/changelog/} (last accessed: 13.01.2026)} and the TikTok Changelog\footnote{\url{https://developers.tiktok.com/doc/changelog} (last accessed: 13.01.2026)}.
The results in this section integrate two complementary types of evidence. The first is quantitative: direct comparisons between our sockpuppet-collected PIE and the data retrievable through Research APIs, yielding concrete coverage and metadata completeness figures. The second is contextual: platform documentation, API specifications, prior literature, and our own experience as vetted researchers navigating both Research APIs' controlled environments, which provide the necessary context to interpret \textit{why} specific content or metadata is missing. We combine both types of evidence because the quantitative audit alone cannot explain the mechanisms behind observed data loss---platforms do not provide error codes or reasons for missing content---while the contextual evidence alone cannot establish the empirical magnitude of the problem. Our typology of access restriction mechanisms (see Sections~\ref{sec:scopesnarrowing}-\ref{sec:operationalrestrictions}) emerged from this combination: We observed systematic patterns of data loss in the audit, then drew on platform documentation and prior work to identify and categorize the underlying mechanisms. Sections~\ref{sec:scopesnarrowing}--\ref{sec:operationalrestrictions} are therefore structured around these mechanism categories, with each subsection presenting the relevant quantitative findings alongside the contextual evidence needed to interpret them.

Overall, we find that researchers can access around 75\% (TikTok \textit{For You} feed) and 50\% (Instagram \textit{Explore} feed) of the posts shown to users. Among those posts that are \textit{accessible}, researchers can access 17\% (TikTok) and 42\% (Instagram) of the metadata parameters that are transmitted to the user's browser.
To structure our comparison, we distinguish three mechanisms along which data loss occurs: (1) \textit{scope narrowing}, referring to which posts are accessible at all (see Section~\ref{sec:scopesnarrowing}); (2) \textit{metadata stripping}, referring to the contextual information removed for accessible posts (see Section~\ref{sec:Metadata-stripping}); and (3) \textit{operational restrictions}, referring to procedural, technical, and economic constraints on accessing these posts (see Section~\ref{sec:operationalrestrictions}). Each of these mechanisms implements a filter on the data that remains accessible through Research APIs, and the overall observed reduction in accessible data is hence the result of what we term \textit{overlapping filters}: multiple restrictions applied simultaneously, each excluding a different subset of content. The effects of these filters accumulate and, since platforms do not provide reasons for why specific content is unavailable through the APIs, it is not possible to attribute precise percentage losses to individual restrictions. Our analysis therefore focuses on the overall accessible share of content rather than attempting to disaggregate losses by filter type.

\subsection{Data Loss through Scope Narrowing}
\label{sec:scopesnarrowing}

The first filter arises from \textit{scope narrowing}, referring to restrictions on which posts are accessible through Research APIs at all. Scope narrowing excludes entire categories of content from research access, independent of any metadata or operational limitations.
This includes posts that are no longer available on the platform, such as content deleted by users or removed through moderation, as well as posts that remain visible to users in the platform’s PIE but are excluded from Research APIs because platforms unilaterally do not classify them as ``publicly accessible.''

\subsubsection{Account- or Entity-Based Loss}

\begin{figure*}[t]
    \centering
    \includegraphics[width=1.00\linewidth]{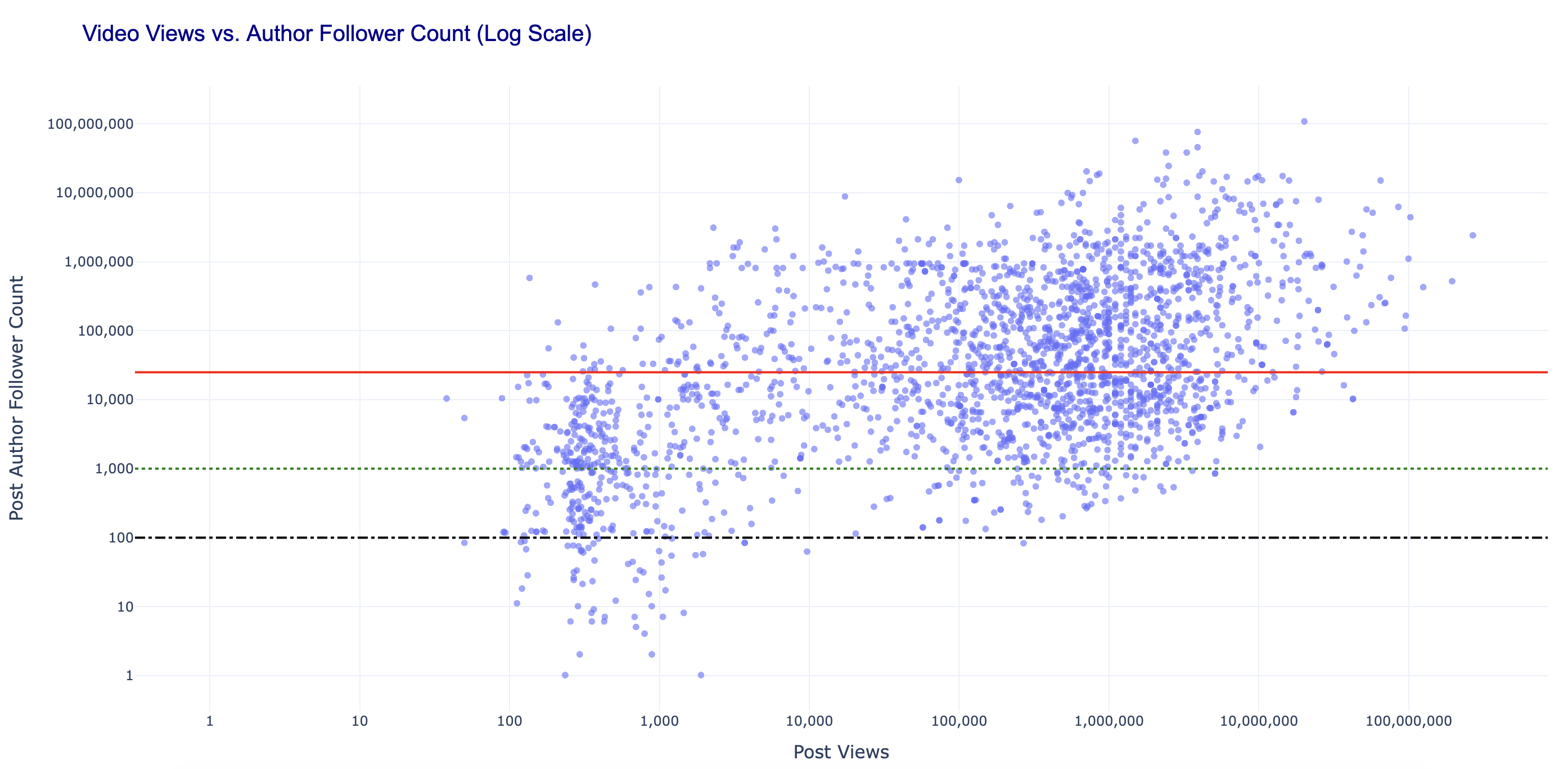}
    \caption{Distribution of follower counts for accounts appearing in the sockpuppet’s Instagram feed. Horizontal reference lines indicate follower-count thresholds applied by Meta’s Content Library at different stages of its implementation: 25{,}000 followers (solid red), 1{,}000 followers (dotted green), and 100 followers (dash–dot black).}
    \Description{A scatter plot on log-log axes showing several thousand 
Instagram posts. The horizontal axis is post views, ranging from 1 to 
over 100 million; the vertical axis is the post author's follower 
count, also ranging from 1 to over 100 million. Each blue dot 
represents one post. The point cloud shows a broad positive 
association: posts with higher view counts tend to come from accounts 
with more followers, but with several orders of magnitude of vertical 
spread at any given view count, and a long lower tail of posts from 
small accounts. Three horizontal reference lines mark follower-count 
thresholds used by Meta's Content Library: a solid red line at 25,000 
followers, a dotted green line at 1,000 followers, and a dash-dot 
black line at 100 followers. A large share of the plotted posts fall 
below the 25,000 line, a visible minority fall below the 1,000 line, 
and a smaller but non-trivial number fall below the 100 line.}
    \label{fig:accountsize}
\end{figure*}

A major source of scope-related data loss arises from account-based eligibility thresholds imposed by platform Research APIs. On Meta platforms, research access has historically been restricted by account size: only public accounts above a minimum follower threshold are included, and only accounts whose follower count surpasses a different, higher, threshold are fully downloadable. These thresholds operationalize a narrow notion of public relevance based on follower count, implicitly equating follower count magnitude with influence. While compelling on first sight, this assumption is problematic, as recommender systems routinely amplify content from small or newly created accounts, enabling posts to reach large audiences regardless of the creator’s follower base (see Figure~\ref{fig:accountsize}). In addition, account-size thresholds systematically exclude high-impact content from national, regional, or niche information environments, where influence is relatively high yet in a local scope, while not globally~\cite{Better_Access}.
Using our Instagram sockpuppet data, we quantify the effect of these thresholds by applying three follower-count cutoffs that correspond to Meta’s current and prior access rules. These thresholds regularly change, and have also changed during our study period: Under the historically applied 25{,}000-follower cutoff, which governed full data access for much of the study period, 49.35\% of user-visible posts originate from accounts that would be entirely excluded from researcher access, leaving only 50.65\% of posts observable via the Meta Content Library. Even under the more permissive 1{,}000-follower threshold introduced in Version~5.0 of the Meta Content Library (November~11,~2024), 11.45\% of posts remain inaccessible. Under the most recent threshold of 100 followers introduced in Version~6.0 (November~10,~2025), about (2.47\%) of posts are excluded. Table~\ref{tab:account_thresholds} reports the share of posts originating from accounts below and above each threshold. 

\begin{table*}[t]
\caption{Impact of Meta account-size thresholds on research-accessible posts in the Instagram sockpuppet feed. Percentages indicate the share of PIE posts excluded or retained under each threshold.}
\label{tab:account_thresholds}
\centering
\begin{tabular}{lrr}
\hline
\textbf{Follower Threshold} & \textbf{Posts Below Threshold} & \textbf{Posts Accessible} \\
\hline
$\geq$ 25{,}000 followers & 49.35\%  & 50.65\%  \\
$\geq$ 1{,}000 followers  & 11.45\%  & 88.55\%  \\
$\geq$ 100 followers      & 2.47\%  & 97.53\%  \\
\hline
\end{tabular}
\end{table*}

Figure~\ref{fig:accountsize} illustrates the underlying follower-count distribution. Notably, a substantial share of posts that reached the sockpuppet through algorithmic recommendation originated from accounts well below the threshold at the time of data collection. Filtering out these accounts therefore removes the majority of the user’s PIE, including viral and politically salient content amplified despite limited account size. Meta has adjusted these thresholds twice over time: lowering them from 25{,}000 followers to 1{,}000 followers (Version~5.0, November~11,~2024), and subsequently to 100 followers (November~10,~2025). While these changes improve coverage, they vividly demonstrate the discretionary nature of platform-defined access criteria. %Crucially, even under the most permissive current threshold, a large portion of user-visible content remains outside the scope of research access, including content that may carry transparency-relevant labels or moderation signals.
TikTok applies a different form of entity-based restriction. Its Research API includes content from public accounts regardless of follower count, but excludes posts created by users under the age of 18. In our dataset, applying this age-based restriction would exclude 0.64\% of posts (5 individual creators) from research access. While quantitatively small in this sample, this exclusion is not trivial from an analytical perspective. Youth participation is unevenly distributed across topics and formats, and even a small number of excluded creators may disproportionately affect research on youth-driven information environments, fringe digital spaces~\cite{Zeng_Schaefer_Oliveira_2022_conspiracy_digital}, and political communication in smaller or regional contexts~\cite{Better_Access}, where the influence of younger creators may be disproportionately high.

\subsubsection{Format-Based Loss}
%ephemeral, live, real-time windows
Ephemeral content is defined as ``communication artifacts, including text, pictures, and videos, that [are designed to] be erased after being display[ed] for a limited period of time.''~\cite{Better_Access}, cf.~\cite{bekavac2026broken}.
Platform reports themselves show the scale: the majority of Instagram users watch \textit{Stories} daily, and on TikTok over 100 million creators have gone \textit{live}, with billions of views. Ephemeral and live formats are widely consumed, influence users in real time, and are often used in moments of heightened political or social relevance (e.g., Elon Musk's livestreams on X with Alice Weidel before the German Federal Election 2025~\cite{Weidel-Musk} and with Donald Trump before the 2024 US elections~\cite{Conger_Mac_2024}).
Meta and TikTok transparency tools and Research APIs operate with a default delay of 24 hours to several days (see Section \ref{sec:real-time-Issues}), which leads to ephemeral and live posts being excluded from Research APIs \textit{by default}. Notably, this is also the case for \textit{Stories} which are made available for longer in accounts under the highlights tab\footnote{\textit{Stories} that are added as \textit{highlights} remain visible until you remove them, even after the original story has disappeared, see \url{https://help.instagram.com/1660923094227526} (last accessed: 13.01.2026)}. Reason is that by the time data would become available to Researcher in those channels, the content has already vanished.
From a PIE perspective, this creates a complete blind spot. The impact of these restrictions is that 0\% of this public available content is accessible through the Research APIs.
It should be noted that our sockpuppet-based data collection methodology captures only feed-based posts delivered through TikTok's \textit{For You} feed and Instagram's \textit{Explore} feed, and was not configured to collect Stories, live streams, or other ephemeral content formats. We therefore cannot quantify from our dataset what proportion of the overall PIE consisted of ephemeral content. However, this limitation does not affect our assessment of Research API coverage for ephemeral formats: since such content exists on the platfrom, we can state that, regardless of the exact overall volume of such content, none of it is accessible through either platform's Research API.

\subsubsection{Moderation and Deletion Loss}
\label{sec:Content-moderation}
%moderation, user deletion, archival
A third filter of data loss arises from content moderation and deletions. Content may disappear from platforms in three principal ways: (1) it may be removed by the platform through moderation or policy enforcement; (2) it may be deleted by users, either at the level of individual posts or entire accounts; or (3) it may be withdrawn from public visibility without being deleted, for example by archiving posts or switching account settings to ``private''~\cite{Better_Access}. While these mechanisms differ in intent and legal implications, they cause a common effect on research access: Content that was once part of the PIE to users becomes inaccessible through research APIs.
This is particularly problematic because the very posts moderated for violating terms of service on misinformation, hate speech, or manipulation—precisely those most relevant for systemic risk research—are the ones missing from researcher’s data access. Even when such posts remain publicly available on the platform, they are not always surfaced through the official research tools. For example, Meta’s Content Library excludes certain categories of content, such as age-restricted material, on the grounds that it is not “publicly accessible”. In addition, both Meta and TikTok impose retrospective deletion obligations on researchers. Meta periodically distributes lists of content ids that have become invalid and requires researchers to delete any corresponding data from their research environments and formally certify compliance, with access suspended for non-compliance.\footnote{\url{https://developers.facebook.com/docs/content-library-and-api/content-library-api/guides/data-deletion/} (last accessed: 13.01.2026)} TikTok similarly requires researchers to refresh their datasets at least every 30 days and delete any content no longer available through the Research API, again with mandatory certification upon request.\footnote{\url{https://www.tiktok.com/legal/page/global/terms-of-service-research-api/en} (last accessed: 13.01.2026)} In both cases, platforms do not provide reasons or API error codes explaining why specific items became unavailable. Content moderation and user-driven deletions thus introduce a temporal dimension of data loss and creates a structural survivorship bias in the data. Research APIs tend to preserve only the “safe” content that remains after moderation, while the systemic risk-relevant material like content flagged, taken down, or archived is erased. As a result, the data accessible to researchers are skewed toward comparatively “safe” content, even though the removed posts may already have reached millions of users.
To quantify this temporal data loss, we re-queried the collected posts in the PIE of the sockpuppet on TikTok at multiple points after initial collection. Table~\ref{tab:tiktok_temporal_loss} reports the share of posts that remained online, were deleted, or became otherwise unavailable (e.g., archived or accounts set to private). Across three observation points, between 17.7\% and 23.3\% of posts were no longer accessible within weeks, indicating rapid and substantial loss of user-visible content.We note that our snapshot-based methodology cannot determine the exact moment at which a post was removed or became unavailable. The three observation points capture cumulative availability at discrete intervals rather than continuous monitoring. Importantly, this loss is not limited to low-reach material. Among posts that became deleted or unavailable, 32.5–42.0\% had already exceeded 100{,}000 plays at the time of observation, and 7.0–9.6\% exceeded one million plays, with maximum observed playcounts reaching up to three million. This shows that temporal data loss also affects high-impact content that has already reached large audiences. 
To assess whether this temporal loss disproportionately affects systemic risk-relevant content, we compared availability rates at the latest observation point between posts classified as politically relevant and those that were not. Among political posts, only 53.57\% remained online, compared to 79.15\% of non-political posts. This indicates that politically relevant content, which is central to systemic risk audits under Article~34(1)(c) DSA, experiences substantially higher removal rates. We did not further inspect the non-political posts that disappeared but these may also be relevant to systemic risks in domains beyond electoral discourse, such as illegal content or the protection of minors.

\begin{table}[t]
\caption{Temporal availability of TikTok posts re-queried after initial data collection. Percentages indicate the share of posts that remained online versus those deleted, not found, or otherwise unavailable.}
\label{tab:tiktok_temporal_loss}
\centering
\begin{tabular}{lrrr}
\hline
\textbf{Observation Date} & \textbf{Online} & \textbf{Deleted} & \textbf{Unavailable} \\
\hline
17 Feb 2025 & 82.27\% & 6.89\% & 10.84\% \\
19 Feb 2025 & 81.51\% & 7.53\% & 10.97\% \\
04 Mar 2025 & 76.66\% & 10.59\% & 12.76\% \\
\hline
\end{tabular}
\end{table}

%These dynamics introduce a pronounced temporal dimension of data loss and produce a structural survivorship bias in research-accessible datasets. Research APIs disproportionately preserve content that remains online and compliant, while posts that were flagged, removed, or archived—often after having reached large audiences—are erased from the PIE available via the Research APIs. As a result, research conducted even days or weeks after an event risks reconstructing a systematically sanitized version of the user-facing information environment, rather than the one displayed in Platforms PIE at the time.
\subsubsection{Real-Time Issues and Missing Posts}
\label{sec:real-time-Issues}

A final source of scope-related data loss arises from delays in how platforms ingest and surface newly created content in their Research APIs. On Meta platforms, newly created Facebook profiles, Instagram accounts or Threads profiles, and content posted by these newly created profiles and accounts may not be included in Meta Content Library for up to 4 days\footnote{\url{https://developers.facebook.com/docs/content-library-and-api/content-library/} (last accessed: 13.01.2026)}. On TikTok, newly uploaded videos can take up to 48 hours before they are indexed and appear in the Research API\footnote{\url{https://developers.tiktok.com/doc/research-api-faq} (last accessed: 13.01.2026)}.
During these ingestion windows, content may already be recommended to users and accumulate substantial reach and engagement. If posts or videos are deleted by their creators or removed through platform moderation within this period, they never enter the Research APIs. As a result, content from short-lived or rapidly moderated accounts—despite being present in the PIE of user during its initial circulation—can be entirely absent from researcher-accessible datasets. Further these delays prevent researchers from observing platforms in real time, even though many systemic risks, particularly those related to elections, coordinated influence campaigns, or fast-moving crises, can unfold within short time windows. They introduce an additional layer of survivorship bias consistent with the patterns described in Section~\ref{sec:Content-moderation}, where only content from the PIE that remains available online is retrievable through official research-access interfaces. 

\subsection{Data Loss through Metadata Stripping}
\label{sec:Metadata-stripping}
The second filter of data loss arises from \textit{metadata stripping}, referring to the systematic removal, omission, or degradation of contextual information associated with posts that remain accessible through Research APIs. While the content itself may be retrievable, large portions of the metadata transmitted to users during normal platform use are either absent from research APIs or provided in inaccurate or outdated form. %This limits researchers’ ability to contextualize the subset of content that remains accessible, including assessing amplification, reach, and platform responses.

\subsubsection{Missing Labels and Flags}

Prior work has demonstrated substantial differences between the metadata exposed through Research APIs and the metadata transmitted to users’ devices during normal platform use. For TikTok, a comparison conducted in July 2023 showed that the Research API exposed 32 parameters, compared to 186 parameters returned by the web API and 845 parameters observed in the Android app \cite{auditing_TikTok}, corresponding to a metadata reduction of approximately 83\% relative to the web interface and over 96\% relative to the mobile application.
Complementing this work, we conduct a similar comparison for the Meta Research API. For a single Instagram post, we observe 236 parameters in the full HTTP response transmitted to the browser, compared to 100 parameters returned by the Meta Content Library API and only 14 parameters exposed through the Content Library user interface (see Figure~\ref{fig:postdata} and Appendix~\ref{app:Full-Parameters}). While all parameters provided through the Meta Content Library are formally documented \footnote{\url{https://developers.facebook.com/docs/content-library-and-api/appendix/data-dictionary} (last accessed: 13.01.2026)}, a large number of parameters transmitted to users are omitted from research access. Notably, many of the omitted parameters correspond to internal labels and flags generated by the platforms themselves.
From the metadata transmitted to users’ devices, these parameters appear to capture signals such as content warnings, fact-checking annotations, monetization eligibility, sharing restrictions, and algorithmic friction or downranking. Such signals are highly relevant for research into systemic risks, as they relate directly to how platforms intervene in content circulation and how recommender systems respond to potentially harmful or sensitive material. However, neither the Meta Content Library nor the TikTok Research API exposes these fields. As a result, researchers are unable to directly observe whether content was flagged, downranked, monetized, or subjected to other platform interventions, even when the underlying post itself remains accessible.
For example, fields suggestive of AI-generated content detection or sharing friction appear in the browser-delivered metadata but are absent from Research API responses. Without access to such contextual signals, researchers cannot assess whether users were informed, whether distribution was algorithmically constrained, or how platform mitigation measures interacted with content visibility. This omission effectively removes the observable traces of platform governance from researcher-accessible data.
Across both platforms, we identify a set of such internal labels that—based on their naming and apparent function—are plausibly relevant for studying transparency, amplification, and moderation dynamics (Table~\ref{tab:data-labels}). These fields are not documented in official API references, and their precise semantics are not publicly specified, therefore our interpretation is necessarily inferential and based on observable naming conventions and their apparent relation to known platform features and interventions.
While not all parameters transmitted to the browser are necessarily meaningful for research question, platforms do not disclose which metadata fields are analytically relevant or how they should be interpreted. In the absence of such guidance, and given that these fields are systematically delivered to users’ devices as part of normal platform operation, we include them in our assessment of metadata completeness as part of the observable PIE.

\subsubsection{Inaccurate and Outdated Data}

In addition to missing metadata, research-access interfaces often provide engagement statistics that are delayed or based on archived values rather than current platform states. TikTok notes that its Research API does not always return live engagement metrics. While user-level endpoints retrieve online data for individual accounts, the video query API relies on archived datasets, meaning that "statistics such as view counts and follower counts may lag behind live values by up to ten days."\footnote{\url{https://developers.tiktok.com/doc/research-api-faq} (last accessed: 13.01.2026)}
Meta’s Content Library similarly provides limited guarantees regarding the timeliness of engagement data. According to platform documentation, view counts for posts created within the last 180 days are refreshed approximately every 24 hours only if the post has accumulated more than ten views during that period; otherwise, view counts are updated every five to seven days. For posts older than 180 days, view counts are refreshed every five to seven days regardless of activity. Meta does not publicly document the update frequency or methodology for other engagement metrics, such as comments or shares.\footnote{\url{https://developers.facebook.com/docs/content-library-and-api/content-library/} (last accessed: 13.01.2026)}
As a consequence, engagement metrics retrieved through Research APIs may substantially diverge from the values visible to users at the same point in time. Prior work has shown that TikTok’s research-access metrics%—including \texttt{view\_count}, \texttt{comment\_count}, \texttt{like\_count}, and \texttt{share\_count}—
are often significantly lower than live values observed on the platform \cite{Pearson17022025}. Such discrepancies hinder the study of algorithmic amplification dynamics, early diffusion processes, and network effects, where precise timing and magnitude of engagement are critical for understanding how content gains visibility and influence.

\subsection{Data Loss Through Operational Access Restrictions}
\label{sec:operationalrestrictions}

Beyond limitations in data scope and metadata, research access to the PIE is further constrained by the operational restrictions under which data can be requested and analyzed from Research APIs. These constraints do not alter which content is nominally accessible, but shape how much of it can be retrieved in practice, at what temporal resolution, and under what technical conditions.
As described in Section~\ref{sec:data-collection-apis}, platforms provide research access to the Research APIs primarily through controlled interfaces, including virtual data enclaves, and virtual compute environments. Through these infrastructures, platforms impose limits on query volume, rate, analysis tooling, data export, and, in some cases, financial cost. Together, these operational constraints function as an additional filtering layer that restricts the effective observability of the PIE for researchers.

\subsubsection{Request Limits}
Request limits imposed by platform Research APIs constrain how much data researchers can retrieve within a given time window, directly shaping the scale, coverage, and temporal resolution of empirical analysis. The Meta Content Library imposes a limit of 1{,}000 queries per rolling seven-day window. In addition, researchers may retrieve a maximum of 500{,}000 data records per seven-day period, with a separate cap of 500{,}000 comment records. Once either limit is reached, further requests are blocked until the rolling window resets. The TikTok Research API applies a different regime, allowing up to 1{,}000 requests per day and a maximum of 100{,}000 records per day across its APIs, with individual video and comment endpoints returning up to 100 records per request.
To contextualize these limits, our TikTok and Instagram sockpuppet accounts consumed on average 200–300 posts per day during browsing sessions of approximately one hour. This level of activity is broadly consistent with existing estimates of typical daily use, which suggest that users spend on the order of 30–60 minutes per day on Instagram and TikTok, respectively, with substantial variation across users and contexts \cite{AlfonsoFuertes2023Instagram,zannettou2024analyzinguserengagementtiktoks}. Under the current request limits, researchers would therefore be unable to monitor more than a small number of user PIEs per day—and in some cases, only a handful per week. %Sustained or large-scale sampling of multiple user feeds, particularly during time-sensitive periods such as elections, is not feasible within these constraints.
%In practice, query budgets can be exhausted rapidly even when monitoring a single user’s PIE. A small number of queries with high predicted result counts may be sufficient to reach the cap. 
Moreover, if recommended posts originate from high-engagement accounts—such as celebrities or major media outlets with millions of interactions—retrieving associated comments for only a few such posts can be enough to exhaust Meta’s entire 500{,}000-comment quota. In our sockpuppet data, the ten most commented posts each received between approximately 240{,}000 and 3{,}200{,}000 comments, meaning that collecting comments for just few of such posts would exceed the weekly limit.

\subsubsection{Analysis and Use Constraints}
\label{sec:analysis_limits}

Beyond limits on query volume, platforms impose restrictions on how research-accessible data may be analyzed, combined, and disseminated. For both Meta and TikTok, research access is provided through controlled interfaces (See Section~\ref{sec:data-collection-apis}) and contractual regimes, so called Restricted Data Use Agreement, that significantly constrain permissible analytical practices.\footnote{See \url{https://somar.atlassian.net/wiki/spaces/somardocs/pages/223412225/Restricted+Data+Use+Agreement} (last accessed: 13.01.2026) \& \url{https://www.tiktok.com/legal/page/global/terms-of-service-research-api/en} (last accessed: 13.01.2026)}
In practice, researchers are required to conduct analyses within platform-controlled or approved environments. Analysis code often requires prior approval, software libraries are restricted by default, and intermediate or final outputs are subject to review before export. In addition, researchers are prohibited from linking platform data with external datasets without explicit authorization: Meta explicitly forbids such linkage absent prior written approval, while TikTok’s research terms prohibit combining research data with external information at the user or device level and restrict use beyond the approved research scope.
These analysis and use constraints limit methodological flexibility and reproducibility. Even when content and metadata are nominally accessible, such restrictions shape what questions can be asked and which methods can be applied. %As a result, they constitute a further operational layer of data loss that affects the effective observability of the public information environment.

%\subsubsection{Access Costs}

%A further restriction arises from the introduction of financial barriers. Until November 2025, access to the Meta Content Library API was only possible through SOMAR, Meta’s designated intermediary. SOMAR remains the only independent environment that is accessible by researchers, as opposed to Meta's web-based interface and Meta's API. Beginning in January 2026, SOMAR will charge researchers a USD~1{,}000 setup fee for new projects and USD~371 per month for continued use of its Virtual Data Enclave (VDE). Existing teams will also incur the monthly fee.

%While the DSA frames researcher access as a compliance obligation—platforms “\textit{shall give access}” to publicly accessible data necessary to study systemic risks, and this obligation is understood as entailing access free of charge \cite{WilmanDSA}, the introduction of recurring charges effectively narrows the pool of researchers who can sustain continuous monitoring. In practice, these fees function as an additional filter on who can access Article~40(12) data.

% Commentaries agree that access should be free of charge. Meta provides options to access data free of charge. The SOMAR option is the only one that is available for a fee. My point here is that access is free of charge -- unless a researcher decides on using the SOMAR environment.

\section{Discussion}
\label{sec:discussion}

This study presents a systematic audit of how well platform Research APIs represent the PIE. Comparing user-visible feeds with researcher-accessible data, we find that Research APIs in both investigated cases---for the TikTok and Instagram services---expose only a partial, delayed, and selectively filtered subset of the respective platforms' PIE that is visible to users on the platform. Data loss occurs through multiple mechanisms---including scope narrowing, metadata stripping, and operational constraints---which together produce a structurally incomplete view of the PIE. These limitations are not isolated implementation issues but reflect platform-defined access regimes. As a result, the PIE accessible through Research APIs diverges systematically from the one experienced by users on the investigated platforms. In the following, we discuss three specifically problematic implications of this before providing first proposals of how DSA provisions should be amended based on our findings.
\paragraph{Implications for} systemic risk research
These access constraints have direct implications for research on systemic risks. Because Research APIs disproportionately exclude short-lived, moderated, ephemeral, and rapidly amplified content, the resulting datasets exhibit strong survivorship bias. Content that is removed, downranked, or otherwise intervened with---which often happens for the very reason that it poses risks related to misinformation, manipulation, or harm---is less likely to be preserved in researcher-accessible data. The systematic, structural, underreporting of such content that we document with our study likely leads to the underreporting of content that is relevant for systemic risk in investigations that make use of Research APIs.
\paragraph{Implications for} metadata loss
This problem is further compounded by the metadata loss that our study documents. Missing or outdated engagement metrics, absent moderation labels, and stripped contextual signals limit researchers’ ability to assess amplification dynamics, coordination patterns, and platform responses. Even when posts themselves remain accessible, the absence of these signals makes it difficult to evaluate exposure, impact, or mitigation. As a result, research relying exclusively on Research APIs risks systematically underestimating both the prevalence and severity of platform-mediated harms.
\paragraph{Implications for} methodological scientific investigation of platforms
Our findings raise principled concerns about the research replicability and the feasibility of valid longitudinal research methods in any project that uses Research APIs as a main data source. Because research-accessible datasets change over time due to content deletions, moderation actions, ingestion delays, and mandatory data refresh requirements, the same analysis conducted at different points in time may yield substantially different results---even when applied to the same event or period. This temporal instability undermines core scientific practices. Researchers cannot reliably reconstruct past information environments, validate prior findings, or compare results across studies conducted at different times. Over longer horizons, Research APIs increasingly reflect a curated PIE of content that has survived moderation and platform-defined retention policies, rather than the one users originally encountered. 
\paragraph{Policy recommendations}
The systematic omission of large portions of a platform's PIE constrains the evidence available for assessing whether platforms are adequately addressing systemic risks. This creates a tension between the \textit{formal availability} of data access mechanisms and their \textit{actual value} for effective oversight: Accountability frameworks that rely on adequate access of researchers to platform data risk becoming procedurally satisfied but substantively weakened. We suggest that current research-access implementations would benefit from closer alignment with the technical reality of implemented platforms as well as with the underlying goal of enabling effective scrutiny of platform-mediated risks. To realign practice with the intent and spirit of Article~40(12), we propose three amendments:

\begin{enumerate}
    \item \textbf{Clarification of the Scope of "Publicly Accessible" Content:} All posts in the PIE of the platform---regardless of account size, format, or subsequent removal---should be included in research access. Any content delivered to out-of-network users (e.g., via Explore, Reels, For You, or Search) and that is visible to all users (and, potentially, non-users) should be considered ``publicly accessible''. The definition of ``publicly accessible'' should hence be substantially broadened in light of Article 40’s core purpose of enabling systemic risk scrutiny, the DSA’s broader framework for external oversight, and the regulation's recognition of mechanisms for balancing competing interests.

    \item \textbf{Obligation to Provide Full Contextual Metadata:} To mitigate the problem of missing contextual metadata around user-displayed content, we propose that Article 40 be amended to mandate that VLOPs should provide, through their Research APIs, the full contextual information that is transmitted to user interface devices while a user interacts with the platform. This includes the metadata discussed in Section~\ref{sec:Metadata-stripping} and extends to all metadata necessary to interpret how content is presented, ranked, and perceived by users, including user identifiers, engagement metrics, content performance signals, and internal system labels or classifications. 

    \item \textbf{Requirement of Operationally Effective Research Access:} Hindrances that materially impair the feasibility of approved research projects, such as the operational access constraints that are documented in this study, should not be permitted in Research APIs. The DSA should be amended to state that, whenever a researcher receives access via a Research API under Article 40, this access must be operationally fit to support the planned research, meaning that technical limitations, request caps, latency constraints, or other access barriers may not render the approved research impracticable in practice. This does not preclude reasonable safeguards, but should require that platforms ensure that Research APIs function as effective instruments for the investigation of systemic risks, rather than as purely formal or symbolic compliance mechanisms.

    \item \textbf{Permission for Independent Collection of Public Data:} The three suggestions above all induce balancing of competing interests (e.g., with respect to request limits vs. practical feasibility of a research project). We propose that this could be mitigated by explicitly permitting researchers to scrape publicly accessible data. This could be implemented to permit validation of platform-provided datasets and for replicability while simplifying the compliance framework for VLOPs. Reasonable safeguards for VLOPs may be applied, and could include IP address whitelisting of approved researchers/research projects or usage of special registered accounts. This very study, in fact, highlights the necessity of such independent collection. We provide a more detailed discussion of the interaction between our methodology and current platform research terms in Appendix~\ref{app:legal-compliance}.

\end{enumerate}

\subsection{Limitations}

Our empirical analysis focuses on politically oriented sockpuppets during electoral periods. This choice is intentional, as election-related content exemplifies the high-stakes information environments central to the DSA’s definition of systemic risks, particularly those affecting civic discourse and electoral processes. This focus, however, may introduce bias. Content moderation and enforcement often intensify during politically salient periods, as platforms prioritize election integrity and misinformation mitigation, potentially altering content availability compared to non-electoral contexts. At the same time, other systemic-risk domains may exhibit equal or greater gaps in research access. For instance, research on risks to minors may face higher exclusion rates due to age-based restrictions. Accordingly, both the magnitude and structure of data loss are likely to vary across different PIE context.
Our analysis makes use of sockpuppet accounts to obtain the actual user-perceived PIE of a platform. We acknowledge that there are research contexts where the use of a sockpuppet undermines the representativeness of the research results \textit{in principle}---since they are, after all, no human users, since they might behave differently from human users, and since the platform might notice sockpuppet activity and treat such accounts differently. However, we argue that the use of sockpuppets does not create any such limitations for the present study. This is because the sockpuppet accounts are merely used to find out the differential between the user-conveyed vs. the researcher-conveyed platform PIE, rather than to replicate the specific behavior of a specific user. Further, Research API access is independent of whether the researcher also runs a sockpuppet-based investigation of the platform (since this cannot be known by the platform).% What remains is a remote possibility that the platforms detected our sockpuppets and then configured their Research APIs to selectively provide less data about sockpuppet activity compared to verified human accounts. However, our comparison with data obtained for human accounts via the Research APIs do not provide any evidence for this.

\section{Conclusions}

Even though access to platform Research APIs is granted only after extensive vetting and is mediated through secured data enclaves and controlled compute environments, the resulting datasets provide only a partial and distorted view of platforms’ PIE. Across multiple mechanisms, substantial portions of user-visible content and contextual metadata are missing, delayed, or selectively filtered. The cumulative effect of these gaps is pronounced: the PIE accessible to researchers is shaped by structural survivorship bias, lacks key metadata needed to interpret platform behavior, and is further constrained by strict rate limits and operational restrictions. This raises a fundamental question as to whether such a constrained view of the PIE can meaningfully support independent auditing and accountability under the DSA. Users as well as search engines have broader access to ``publicly accessible'' material simply by browsing and indexing\footnote{\url{https://www.facebook.com/privacy/policy?annotations[0]=3.ex.5-SearchEnginesYouCan&subpage=3.subpage.3-PublicContentWhatContent} (last accessed: 13.01.2026)} the platform. Commercial scraping firms, which operate at scale, routinely extract richer datasets than academics are allowed to access (cf.~\cite{Miles2025Dashboards}). And even platforms' own AI chatbots and LLM-based tools have more comprehensive access to publicly visible content\footnote{\url{https://www.facebook.com/privacy/genai/} (last accessed: 13.01.2026) ; furthermore, cf. \url{https://micahflee.com/grok-will-share-detailed-responses-from-xs-api-but-alas-it-cant-access-dms-and-likes/} (last accessed: 13.01.2026)} than vetted researchers acting under a legal mandate. The core issue therefore is not technical, but stems from interpretation of DSA provisions by platforms---for instance towards what is "publicly accessible data"---and from platforms creating their own practical access barriers, which points at regulatory loopholes in the DSA. 
Based on our findings, we therefore argue that current research-access implementations fall short of the DSA’s intended oversight function and outline concrete directions for reform. Without such changes, research relying on official VLOP Research APIs will remain systematically biased and insufficiently representative of the user-facing PIE. As a result, the empirical evidence required for effective scrutiny of systemic risks is withheld, leaving a central pillar of the DSA’s accountability framework structurally weakened.

%%
%% The next two lines define the bibliography style to be used, and
%% the bibliography file.
\bibliographystyle{ACM-Reference-Format}
\bibliography{references.bib}

\appendix

\section{Potential Systemic Risk Related Parameters Transmitted via Instagram API Call}
\label{app:Full-Parameters}

\begin{table*}[h]
\caption{Instagram Platform internal labels relevant to transparency, amplification, and moderation signals}
\label{tab:data-labels}
\centering
\begin{tabular}{|p{6.1cm}|p{8.4cm}|}
\hline
\textbf{Data Label Key Name} & \textbf{Potential Relation to Systemic Risks Research} \\
\hline
\verb|boost_unavailable_identifier| & Identifies a blocked boost instance (paid amplification denial). \\ \hline
\verb|boost_unavailable_reason| & Reason boosting is denied; indicates ad policy enforcement. \\ \hline
\verb|boost_unavailable_reason_v2| & Updated denial reason; tracks evolving enforcement logic. \\ \hline
\verb|should_request_ads| & Post eligible/queued for ads; impacts algorithmic reach incentives. \\ \hline
\verb|is_paid_partnership| & Branded content flag; transparency for commercial influence. \\ \hline
\verb|is_eligible_content_for_post_roll_ad| & Monetization eligibility; shapes recommendation incentives. \\ \hline
\verb|related_ads_pivots_media_info| & Links post to ad pivots; signals amplification pathways. \\ \hline
\verb|commerce_integrity_review_decision| & Commercial moderation outcome; integrity checks for promotions. \\ \hline
\verb|integrity_review_decision| & Core moderation verdict; content safety intervention signal. \\ \hline
\verb|sharing_friction_info| & Indicates share friction; throttles virality for risky content. \\ \hline
\verb|should_have_sharing_friction| & Explicit friction trigger tied to risk score. \\ \hline
\verb|community_notes_info| & Availability of community corrections; misinformation control. \\ \hline
\verb|comment_inform_treatment| & Warning prompts before commenting; harm mitigation. \\ \hline
\verb|gen_ai_detection_method| & AI-origin detection mechanism; provenance transparency. \\ \hline
\verb|has_high_risk_gen_ai_inform_treatment| & High-risk AI content flagged; stricter moderation. \\ \hline
\verb|gen_ai_chat_with_ai_cta_info| & AI assistant overlays; system-driven engagement nudges. \\ \hline
\verb|is_quiet_post| & Reduced distribution/notifications; shadow-throttling signal. \\ \hline
\verb|like_and_view_counts_disabled| & Hidden engagement metrics; safety and harassment mitigation. \\ \hline
\verb|ig_media_sharing_disabled| & Sharing disabled; prevents further distribution. \\ \hline
\verb|hide_view_all_comment_entrypoint| & Restricts comment exploration; limits harmful reply spread. \\ \hline
\verb|can_view_more_preview_comments| & Controls comment expansion; containment lever. \\ \hline
\verb|report_info.has_viewer_submitted_report| & Community reporting activity; moderation trigger. \\ \hline
\verb|is_open_to_public_submission| & Public content submissions allowed; raises manipulation risk. \\ \hline
\verb|media_notes| & Platform annotations on media; quality and safety signals. \\ \hline
\verb|meta_ai_suggested_prompts| & Platform-generated suggestions; automated engagement nudging. \\ \hline
\end{tabular}

\end{table*}

\clearpage

\section{Post-Matching Pipeline for Research API Coverage Calculation}
\label{app:data-pipeline}

This Section describes the procedures used to determine which posts from the sockpuppet-collected public information environment (PIE) were retrievable through the platforms' official Research APIs. Because the two platforms impose different identifier regimes and query constraints, separate matching strategies were required.
\subsection{TikTok Research API}

For TikTok, matching follows the approach established in prior audit work \cite{entrenaserrano2025tiktoksresearchapiproblems, Pearson17022025}. Each TikTok post URL contains a native video\_id that serves as a unique, platform-wide identifier. The TikTok Research API accepts this identifier as a direct query parameter, enabling deterministic one-to-one matching between sockpuppet-collected posts and API-accessible records. For each post in the sockpuppet PIE, we queried the Research API using the corresponding video\_id. A post was counted as retrievable if the API returned a valid response containing the video's metadata. Posts for which the API returned no result or an error were classified as not retrievable. Queries were conducted at multiple points between March and June 2025.
\subsection{Meta Content Library}

Matching posts between the Instagram sockpuppet PIE and the Meta Content Library is substantially more complex, because Meta generates privacy-preserving identifiers for both creators and posts that do not correspond to the platform's native identifiers. The Content Library does not support queries via Instagram post URLs, native post IDs (PK-IDs), or Instagram usernames directly. This limitation has been noted in prior work as a barrier to cross-referencing publicly obtained datasets with Content Library records~\cite{2024-edmo}. We therefore implemented a multi-stage matching pipeline:

\begin{enumerate}
    \item \textbf{Creator-level matching.} For each post in the sockpuppet PIE, we extracted the creator's Instagram username and constructed the corresponding account URL (\url{https://www.instagram.com/{username}/}). The Meta Content Library Interface supports bulk upload of account URLs via CSV files, which are then resolved against the Library's internal creator database. This step produces a \textit{Producer List}---a reusable, queryable set of matched creators within the Content Library environment. Creators whose URLs could not be resolved (e.g., because the account fell below Meta's follower threshold, had been set to private, or was otherwise excluded from the Content Library) were classified as \textit{not accessible}. All posts from inaccessible creators were accordingly classified as \textit{not retrievable} through the API, contributing to the account-level component of scope narrowing documented in Section~\ref{sec:data-collection-apis}.

    \item \textbf{Post-level matching within accessible creators.} For the subset of posts originating from creators present in the Producer List, we conducted post-level searches within the Content Library Interface. Because direct post-ID lookups are not supported, searches were performed using post captions (descriptions) as query strings, restricted to the matched Producer List and bounded by the relevant time window. Specifically, for each sockpuppet-collected post with a unique caption, we constructed a Content Library search URL encoding the caption text, the Producer List reference, and a date range encompassing the post's creation date. A post was counted as \textit{retrievable} if the search returned a matching result from the correct creator within the expected time window.

    \item \textbf{Edge-case handling.} Two categories of posts required additional manual verification:
    \begin{itemize}
        \item \textit{Duplicate captions:} Six distinct captions appeared across multiple posts from different creators, affecting a total of 35 posts. Because caption-based search could not disambiguate these, we verified each post individually by inspecting the creator's full post history within the Content Library.
        \item \textit{Empty captions:} 79 posts had no caption text, making caption-based search infeasible. For these, we retrieved the full set of posts from the respective creator within the relevant time window and manually identified whether the target post was present based on creation date and available metadata.
    \end{itemize}
\end{enumerate}

\paragraph{Replicability considerations}

The Producer List used in this study is accessible to any researcher with Meta Content Library access at \url{https://www.facebook.com/transparency-tools/content-library/producer-lists/509889644923339/}. However, exact replication of our coverage results is inherently limited by the temporal dynamics documented in this paper: creators may change their account visibility (e.g., switching to private), delete posts, or fall below updated follower thresholds between the time of our analysis and any replication attempt. This instability is itself a manifestation of the overlapping filters described in Section \ref{sec:overlapping-filters}. We make our full dataset available to vetted researchers upon request for cross-verification purposes. Due to the presence of personal and politically sensitive content in the sockpuppet PIE, the dataset is not released publicly but is shared under appropriate data protection safeguards with other vetted researchers.

\subsection{Temporal Alignment}

Sockpuppet data collection covered the periods July 16–November 6, 2024 (TikTok, U.S. presidential election) and January 8–February 25, 2025 (Instagram, German federal election). Research API queries for both platforms were conducted at multiple points between March and June 2025. The time gap between initial PIE collection and API querying means that the coverage rates reported in Section 4 incorporate any content that became unavailable in the interim due to deletions, moderation actions, or account-visibility changes—consistent with the temporal data loss patterns documented in Section \ref{sec:scopesnarrowing}

\section{Sockpuppet Configuration and Primer Prompts}
\label{app:Sockpuppets}

This Section provides additional detail on the sockpuppet accounts used in this study and the primer prompts that guided their interaction with platform content. Both sockpuppets were operated using the SOAP framework~\cite{Bekavac-TORS-SOAP}, an open-source tool specifically designed for auditing systemic risks in personalized recommender systems on VLOPs. SOAP was developed with the regulatory objectives of the DSA in mind, in particular the need for independent scrutiny of systemic risks under Article~40. It enables controlled interaction with social media platforms through sockpuppet accounts, captures the full HTTP response payload transmitted to the user's browser, and integrates a multimodal LLM (Gemini 1.5 Flash) for automated deductive coding of encountered content. SOAP accesses only publicly accessible data from platform feeds and does not interact with non-public backend systems or circumvent technical access controls. Its interaction patterns---limited to viewing and liking posts with delays simulating natural scrolling---are deliberately designed to minimize platform impact while remaining within established norms for black-box auditing and scraping-based research under the DSA~\cite{Bekavac-TORS-SOAP}. A detailed discussion of how our use of SOAP interacts with platform terms of service is provided in Appendix~\ref{app:legal-compliance}.

\subsection{Account Configuration}
\label{sec:appendix-account-config}

Table~\ref{tab:sockpuppet-config} summarizes the configuration of the two sockpuppet accounts.
 
\begin{table*}[h]
\centering
\caption{Sockpuppet account configuration.}
\label{tab:sockpuppet-config}
\begin{tabular}{lll}
\toprule
\textbf{Parameter} & \textbf{TikTok Sockpuppet} & \textbf{Instagram Sockpuppet} \\
\midrule
Platform & TikTok & Instagram \\
Feed type & For You & Explore \\
Collection period & 08.01.2025--25.02.2025 & 16.07.2024--06.11.2024 \\
Election context & 2025 German federal election & 2024 U.S.\ presidential election \\
Location & Switzerland & Switzerland \\
Age & Over 18 & Over 18 \\
Gender & Male & Male \\
Followed accounts & German political accounts & U.S.\ political accounts \\
Interaction type & Liking LLM-flagged posts & Liking LLM-flagged posts \\
Session structure & Multiple sessions across days & Multiple sessions across days \\
Political content share & $\approx$9.7\% of feed & $\approx$41\% of feed \\
\bottomrule
\end{tabular}
\end{table*}

Both accounts followed a set of politically relevant public accounts to establish a baseline political interest signal for the platform's recommender system. The TikTok account followed accounts associated with German political parties and political media; the Instagram account followed accounts associated with U.S.\ political figures and media. Beyond this initial configuration, the sockpuppets' feeds evolved organically through the platform's recommendation algorithm in response to the SOAP-driven interactions (liking of flagged posts).
 
\subsection{Session Structure}

Data collection was conducted across multiple discrete sessions on different days throughout each collection period. SOAP was not run continuously but rather in sessions that approximated typical user browsing behavior, with each session navigating the algorithmic feed and collecting all delivered posts and metadata. Delays of 1--10 seconds between interactions were implemented to simulate natural scrolling speed, and sessions were limited to approximately 100 posts each, corresponding to roughly one to two hours of feed navigation~\cite{Bekavac-TORS-SOAP}. In total, the two sockpuppets collected over 4{,}000 posts across more than 200 hours of cumulative feed navigation.

\subsection{Primer Prompts}
\label{sec:appendix-prompt-text}

SOAP uses primer prompts to guide the integrated multimodal LLM in classifying whether encountered posts are relevant to the research topic. Posts classified as relevant are flagged for interaction (liking), which in turn influences the platform's subsequent recommendations. The primer prompts used in this study are reproduced below. Their design follows the validated SOAP methodology, where prompt reliability has been established through intra-rater (Cronbach's Alpha $> 0.97$) and inter-rater (Cohen's Kappa $> 0.73$ for unambiguous topics) validation.
 
\subsubsection{Instagram Sockpuppet: U.S.\ Election Primer Prompt}
 
\begin{quote}
\small
\texttt{You are analyzing Social Media videos related to the US election and US politics from the perspective of a voting US American.}
 
\texttt{Post Description: \{post\_text\}}\\
\texttt{Username: \{creator\_id\}}
 
\texttt{Please answer the following questions:}\\
\texttt{- On a scale of 0 to 5, how much is the video related to the US election, US politics, or the US voters in general? (0 being not at all related and 5 being entirely focused on this theme)}\\
\texttt{- What is your reasoning for the score?}
 
\texttt{If the post is related to the US election or US politics, rate it higher. Videos that feature US politicians, political debates, or legislative updates should also be rated high.}
 
\texttt{Give your answer precisely in the following format:}\\
\texttt{"Score:Reasoning". For example, "5:The video is entirely about positive aspects of left-leaning US politics." Do not say `Score' or `Reasoning' in the answer.}
\end{quote}
 
\subsubsection{TikTok Sockpuppet: German Federal Election Primer Prompt}
 
\begin{quote}
\small
\texttt{Role Description:}\\
\texttt{You are a neutral expert in political content analysis. Your task is to assess TikTok videos to determine whether they present, disseminate, or support the party Alternative for Germany (AfD), its messages, or its politicians in a positive or negative manner.}
 
\texttt{Scope of Analysis:}\\
\texttt{1. Visual Symbolism: Use of AfD colors (especially blue) in an AfD-related context. AfD logos, election posters, or speeches by AfD politicians. Blue heart emojis in noticeable frequency as an indication of AfD sympathy. National symbols (German flag, eagle) only if used in an AfD-related context. Negative Indicators: Ironical use of AfD symbols, edited content that misrepresents AfD figures in a critical way.}
 
\texttt{2. Audio (Music, Spoken Content): Do the song lyrics mention AfD? Common AfD-associated lyrics, such as: ``Es braucht die AfD'' (We need the AfD). ``Blau, blau, blau'' (Blue, blue, blue). Background audio from AfD campaign speeches or politicians. Negative Indicators: Satirical overdubbing, mocking voice-overs, or negative framing of AfD speeches.}
 
\texttt{3. Account Name: Does the name indicate support for the AfD? Use of terms like ``blue'', ``patriot'', ``alternative'', ``Germany'' in a pro-AfD context. Negative Indicators: Account names with phrases like ``NoAfD'' or ``Antifa''.}
 
\texttt{4. Description \& Hashtags: Pro-AfD Slogans: ``Sei schlau, w\"ahl blau'' (Be smart, vote blue). ``Zeit f\"ur Alice'' (Time for Alice). Pro-AfD Hashtags: \#seischlauw\"ahlblau, \#zeitf\"uralice, \#afdwirkt, \#afd. Anti-AfD Hashtags (Indicating rejection of AfD, therefore score = 0): \#fckafd, \#noafd.}
 
\texttt{Evaluation Scale (0 to 5):}\\
\texttt{0: No reference to AfD or clear rejection of AfD (e.g., using \#fckafd or satire).}\\
\texttt{1--2: Neutral mention of AfD (e.g., journalistic reporting) or indirect association.}\\
\texttt{3--4: Clearly positive representation of AfD but not entirely focused on promotion.}\\
\texttt{5: Consistent and direct support or promotion of AfD, its politicians, or messages.}
 
\texttt{Analysis Process:}\\
\texttt{1. Check Context: What are the visual and acoustic elements? Do song lyrics contain AfD-related statements? How are symbols, statements, and hashtags used?}\\
\texttt{2. Classification and Evaluation: Is the AfD represented positively, neutrally, or negatively? Are there clear signs of support or promotion? Mixed Signals: If the video contains both pro- and anti-AfD elements, evaluate the dominant message before assigning a score.}
 
\texttt{Response Format (ALWAYS STRUCTURE YOUR RESPONSE LIKE THIS):}\\
\texttt{"[Score]:[Reasoning]" Do Not use brackets [] in your response.}
 
\texttt{Examples:}\\
\texttt{Interpretation: The video calls for voting for the AfD, features a speech by Alice Weidel, uses blue heart emojis, and includes the lyrics `Es braucht die AfD'. Score: 5}\\
\texttt{Interpretation: The video shares AfD positions on migration but does not explicitly endorse voting for the party. The song lyrics have no AfD reference. Score: 3}\\
\texttt{Interpretation: The video includes the hashtag \#noafd, indicating rejection of the AfD. Score: 1}
 
\texttt{Here the Video Information:}\\
\texttt{Post Description: \{post\_text\}}\\
\texttt{Username: \{creator\_id\}}
\end{quote}
 
\subsection{Content Classification and Filtering}

The primer prompts classify each post on the specified scale. For interaction purposes, posts exceeding a relevance threshold were flagged and liked by the sockpuppet. Importantly, for the Research API comparison reported in this paper, \textit{all} posts collected by the sockpuppets were included in the analysis, regardless of their political relevance classification. The coverage rates reported in Section~\ref{sec:overlapping-filters} therefore reflect the overall accessibility of the full PIE, not only of its political subset.
Non-political content was not filtered out of the dataset prior to the Research API comparison. This design choice ensures that our coverage estimates are representative of the complete user-facing PIE rather than being biased toward a particular content category. The political content percentages reported above (9.7\% for TikTok, 41\% for Instagram) describe the composition of the collected feeds, not the scope of the analysis.

\section{Reflections on Platform Terms of Service and Research Norms}
\label{app:legal-compliance}

Our methodology requires reconstructing the user-visible PIE through sockpuppet accounts and browser-level data capture, and subsequently matching this baseline against Research API outputs. Both steps interact with platform research terms in ways that merit transparent discussion. TikTok's Research Terms (Section~III) restrict access to TikTok content to the official Research Tools and prohibit scraping or manual extraction. Meta's terms similarly prohibit automated data collection outside the Content Library.\footnote{\url{https://www.tiktok.com/legal/page/global/terms-of-service-research-api/en} (last accessed: 13.01.2026); \url{https://transparency.meta.com/researchtools/meta-content-library/} (last accessed: 13.01.2026).} Our sockpuppet-based collection, operated through the SOAP framework~\cite{Bekavac-TORS-SOAP}, might be seen as falling outside these permitted channels. However, independent ground-truth reconstruction is a prerequisite for validating the completeness of platform-provided data---without it, the very question this paper investigates cannot be scientifically answered. This follows established auditing practices in computer science and security research, where controlled observation of system outputs is necessary to evaluate system behavior (e.g.,~\cite{Sandvig2014AuditingA}). 
We recognize that platforms face legitimate obligations to protect user privacy and to prevent malicious data extraction, and that anti-scraping measures play an important role in this regard. However, these concerns do not justify preventing lawful, vetted researchers from accessing publicly visible data necessary for independent scrutiny. As the European Data Protection Supervisor has emphasized, platforms may not “escape accountability [...] on the pretext of safeguarding the rights of others”~\cite{Keller2025PubliclyAccessible}. At the same time, we acknowledge the structural tension platforms face: more robust data access enables stronger external scrutiny, which may reveal non-compliance or systemic risks and expose platforms to regulatory consequences.
Furthermore, both Meta (Section~5(b)(iii) of the Restricted Data Use Agreement) and TikTok (Section~III, Privacy Obligations) restrict combining Research Data with external datasets. Our study necessarily linked Research API outputs with sockpuppet-collected posts to measure coverage and metadata completeness. This linkage was conducted strictly at the level required for this study, and all reported results are aggregated statistics. TikTok additionally requires periodic dataset refresh and deletion of unavailable content, as well as advance sharing of publications; Meta restricts export of raw data from its virtual compute environment. Our study reports exclusively aggregate findings and did not require exporting any data from the environment.

These constraints reveal a structural tension as independent auditing of Research API completeness requires precisely the validation techniques---scraping for ground truth and cross-dataset comparison---that platform terms restrict. This tension is not unique to our study. The European Commission is debating whether scraping should be recognized as a legitimate means of researcher access to publicly accessible platform data\footnote{See \url{https://www.techpolicy.press/what-the-x-fine-reveals-about-data-access-under-article-40-of-the-digital-services-act/} (last accessed: 24.03.2026)}, and our fourth policy recommendation (Section~\ref{sec:discussion}) addresses this directly. In this context, we further refer to the EDMO and European Commission's Status Report on researcher access mechanisms~\cite{Status_Report:_Mechanisms_for_researcher_access2024, 2024-edmo}, which documents similar methodological constraints faced by researchers across multiple platforms. We conducted this study for non-commercial, public-interest research purposes under institutional research oversight. The sockpuppet accounts collected only publicly accessible content delivered during ordinary platform use and did not access non-public backend systems or circumvent technical access controls. We acknowledge the contractual friction but argue that the ability to independently verify the completeness of DSA-mandated data access mechanisms is essential to the regulatory framework's integrity. A research ecosystem in which the adequacy of platform-provided access can only be assessed using platform-provided tools would be structurally incapable of detecting the very shortcomings this paper documents.

\section*{Endmatter}

\subsection*{Generative AI Usage Statement}

The authors used a generative AI tool (ChatGPT, OpenAI, version GPT-5.2) for editorial support, including grammar correction, clarity improvements, stylistic refinement of text written by the authors, and assistance with table formatting and organization. The tool was not used to draft or generate original manuscript text, scientific claims, analyses, results, or interpretations. All intellectual content and responsibility for the manuscript remain with the authors.

\end{document}